# The mass of the dominant particle in a fractal universe

Scott Funkhouser[1] and Nicola Pugno[2]
[1]Department of Physics, the Citadel, 171 Moultrie St. Charleston, SC, 29409
[2]Department of Structural Engineering, Politecnico di Torino, Corso Duca degli Abruzzi 24, 10129 Torino, Italy

ABSTRACT

An empirically validated, phenomenological model relating the parameters of an astronomical body to the stochastic fluctuations of its granular components is generalized in terms of fractal scaling laws. The mass of the particle constituting the preponderance of the mass of a typical galaxy is determined from the generalized model as a function of the fractal dimension. For a fractal dimension between 1 and 3 the mass of the dominant particle in galaxies is, roughly, between the Planck mass and 1eV. If the dimension is near 2 then the fractal model is identical to the original stochastic model, and the mass of the dominant particle must be of order near the nucleon mass. Two additional expressions for the mass of the dominant particle in the universe are obtained from basic quantum considerations and from the existence of a cosmological constant. It follows that the fractal dimension 2 is favored and that the mass of the dominant particle is proportional to sixth root of the cosmological constant and of order near the nucleon mass.

*1. The generalized CMSI model and fractal structure*

According to a phenomenological scaling law based on the stochastic motions of the granular components of a large, gravitational body, the characteristic parameters of the observable universe, galaxies, solar systems and the nucleon may be related through a set of self-similar scaling laws. Let there be an astronomical body, denoted by *j*, that is gravitationally bound, or nearly so, and that consists of some very large number $N_{kj}$ of components, denoted by *k*, that are very small with respect to the body that and may be treated as granular. The number $N_{kj}$ should be of the order $M_j/M_k$, where $M_j$ and $M_j$ represent respectively the masses of *j* and *k*. If the dynamics of the astronomical body should be determined statistically by the stochastic fluctuations of the components then it follows that the characteristic relaxation-time $t_j$ of the astronomical body and the characteristic time $t_k$ of the statistical fluctuations associated with any one of the granular components should be, on average, related by

$$t_j \sim t_k N_{kj}^{1/2}. \qquad (1.1)$$

As a consequence of the primary relationship in (1.1), the characteristic radius $R_j$ of the astronomical body should be scaled to the characteristic radius $R_k$ of the component *k* by

$$R_j \sim R_k N_{kj}^{1/2}. \qquad (1.2)$$

Furthermore, the characteristic action $A_j$ of the astronomical body should be related to the action $A_k$ of the granular component according to

$$A_j \sim A_k N_{kj}^{3/2}. \qquad (1.3)$$

The physical model from which the relationships in (1.1), (1.2) and (1.3) were derived is identified here as the "CMSI" model, after the four authors of its seminal presentation in Ref. [1].

Suppose that each component *k* is also a gravitational, astronomical system that consists essentially of a large number $N_{lk}$ of relatively small components *l*. It follows that the relationships in (1.1), (1.2) and (1.3) should relate the parameters of the bodies *k* and *l* in the same manner in which they relate the parameters of the bodies *j* and *k*, respectively. Furthermore, (1.1), (1.2) and (1.3) should relate the parameters of the bodies *j* and *l* in they same manner in which they relate, respectively, *j* and *k* and *k* and *l*. In that manner the scaling laws of the CMSI model are inherently self-similar, and accommodate naturally self-similar hierarchies of structure. The smallest granular component whose stochastic fluctuations could determine the dynamics of any given astronomical system $j$ is some species of microscopic particle, denoted by $m$. Consequently, the scaling laws (1.1), (1.2) and (1.3) relate the parameters of the body $j$ to the microscopic component $m$ in the same manner in which they relate the parameters of the body $j$ to any other astronomical component. The CMSI model establishes therefore scaling relationships among astronomical parameters and the fundamental parameters of particle physics.

Let the subscripts $n$, $s$, $g$ and $u$ denote respectively the nucleon, solar systems, galaxies and the observable universe. Even if the bulk of the masses of galaxies and the observable universe are non-baryonic, the difference between $N_{ng}$ and $M_g/M_n$ would be very small with respect to $N_{ng}$, and the difference between $N_{nu}$ and $M_u/M_n$ would be very small with respect to $N_{nu}$. Similarly, the numbers $N_{sg}$, $N_{su}$ and $N_{gu}$ would differ from their respective approximations $M_g/M_s$, $M_u/M_s$ and $M_u/M_g$ by relatively small amounts if some non-baryonic particle dominates the masses of the observable universe and galaxies. The nucleon represents presumably the fundamental granular component of solar systems and, even in the presence of a significant number of some species of non-baryonic particle, represents with a relatively small error a fundamental granular component of galaxies and of the observable universe. Stars similarly represent, with a relatively small error, a granular component of galaxies and the observable universe. Galaxies constitute effectively a granular component of the observable universe. The observable universe, galaxies, solar systems and nucleons constitute therefore a hierarchy of structures that should be subject to the scaling laws associated with the CMSI model. The characteristic parameters of the nucleon, solar systems, galaxies and the observable universe are empirically consistent with the mutual relationships in (1.1), (1.2) and (1.3), as well as the other predictions of the CMSI model, constituting a strong validation of the model. [1]

According to the dark matter paradigm, galaxies and the observable universe are composed primarily of some yet-unidentified, non-baryonic particle. It follows from the CMSI model that the stochastic fluctuations of the putative dark matter particle should determine the dynamical parameters of galaxies and of the observable universe. The scaling laws in (1.1), (1.2) and (1.3) should relate therefore the parameters of the particle to the parameters of galaxies and the observable universe, supplanting the presumably approximate relationships that followed from treating the masses of galaxies and the observable universe as primarily baryonic. The purpose of this present work is to employ a generalized version of the CMSI model in order to obtain limits on the mass of the particle that constitutes the bulk of the masses of galaxies and the observable universe. (The particle that constitutes the bulk of the mass of an astronomical body is called here the "dominant particle" of that body.) The present analysis is based on no presumptions about the identity or mass of the dominant particle, and allows for the possibility that the nucleon represents the dominant particle of galaxies and the observable universe.

It is convenient to express the relationships (1.1), (1.2) and (1.3) in an alternative but equivalent form. Let there be some hierarchy of astronomical structure that is subject to the CMSI model and that contains some unspecified number of levels. In general, the number $N_{ba}$ of some component $b$ that constitutes an aggregate astronomical body, denoted by $a$, is of the order $M_a/M_b$, where $M_a$ and $M_b$ represent respectively the masses of $a$ and $b$. The relationships in (1.1), (1.2) and (1.3) may be therefore expressed as

$$\frac{t_i}{M_i^{1/2}} \sim o, \tag{1.4}$$

$$\frac{R_i}{M_i^{1/2}} \sim q, \tag{1.5}$$

and

$$\frac{A_i}{M_i R_i} \sim f, \tag{1.6}$$

for any body *i* within the hierarchy, where *o*, *q* and *f* are constants. Note that (1.6) follows from (1.3) and (1.5).

The scaling laws in (1.5) and (1.6) may be interpreted more generally expressions of fundamental scaling relationships of fractal structure [2]. If a quantity $M_1$ of mass is arranged in the manner of a fractal hierarchy then there are associated with the mass some number of discrete scales of structure where, for any scale *i*, $M_1$ is represented effectively by an ensemble of bodies, each denoted also by *i* and having a mass $M_i$ and a characteristic radius $R_i$. In a fractal hierarchy whose characteristic dimension is *D* the average density $\rho_j \sim M_j R_j^{-3}$ of matter within the bodies populating some scale *j* is related to the density $\rho_k$ associated with the bodies on some other scale *k* according to [3]

$$\frac{\rho_j}{R_j^{D-3}} \sim \frac{\rho_k}{R_k^{D-3}}. \tag{1.7}$$

The relationship in (1.7) may be expressed as

$$\frac{M_i}{R_i^D} \sim q_D, \tag{1.8}$$

for all scales *i*, where $q_D$ is a constant for a given dimension *D*. The relationship in (1.8) is equivalent to (1.5) if *D*=2, and the constant $q_2$ must be therefore equal to *q*, defined in (1.5).

The characteristic quantities of action $A_j$ and $A_j$ associated with the bodies *j* and *k*, populating any two scales *j* and *k* within a fractal hierarchy, are related according to [2]

$$A_j \sim A_k \left(\frac{M_j}{M_k}\right)^{(D+1)/D}. \tag{1.9}$$

It follows from (1.8) that the relationship in (1.9) may be expressed also as

$$\frac{A_i}{M_i R_i} \sim f, \tag{1.10}$$

for all scales *i*, where *f* is identical to the constant *f* defined in (1.6). The relationship in (1.10) is equivalent to (1.6) since (1.10) is independent of the fractal dimension *D*.

*2. The mass of the dominant particle in galaxies*
Suppose, according to the CMSI model, that the characteristic dynamical parameters of an astronomical body *i* emerge from the stochastic fluctuations of its granular components. Let the body consists primarily of some large number $N_{xi}$ of microscopic particles *x*, which may be either baryonic or non-baryonic, and which represent the smallest granular component. Furthermore, consider the scenario in which the CMSI model is generalized, according to (1.7) – (1.10), in terms of a fractal dimension that may not be 2. Let the fractal dimension relating the dominant particle to the parameters of the astronomical body be $D_{xi}$. The analysis of this Section is entirely consistent with the CMSI model if $D_{xi}=2$, but for the sake of this present investigation the dimension will be left as an unspecified parameter. No other fractal scaling laws other then (1.7) – (1.10) are presumed or necessary.

The characteristic action of any fundamental particle must be of the order the Planck quantum $\hbar \equiv h/(2\pi)$. It follows from (1.9) and $A_x \sim \hbar$ that the mass $M_x$ of the dominant particle $x$ of the body $i$ should be given by $m_i(D_{xi})$, where

$$m_i(D) \equiv M_i \left( \frac{\hbar}{A_i} \right)^{\frac{D}{D+1}}. \tag{2.1}$$

Let the characteristic radius $R_g$ of a typical galaxy be of the order $10^{20}$m, and let the mass $M_g$ be given by the apparent Newtonian dynamical mass of a typical galaxy, which is of the order $10^{43}$kg. (This present investigation is intended to be reliable only to within an order of magnitude, and is insensitive to coefficients of the order $10^0$. However, in order to be consistent, each parameter that is specified by only an integer power of 10 is multiplied by the representative, average coefficient 5.5 in the calculations.) The characteristic action $A_g$ of a galaxy is of the order $U_g t_g$, where $U_g$ is the magnitude of the characteristic gravitational potential energy and $t_g$ is the characteristic relaxation time or transit time. The term $U_g$ is given by $GM_g^2 R_g^{-1}$, which is of the order $10^{56}$J, where $G$ is the Newtonian gravitational coupling. The characteristic transit time is empirically of the order $10^{15}$s. The term $A_g$ is therefore of order near $10^{72}$Js. It follows from (1.10) and the nominal galactic parameters specified here that the constant $f$ must be of the order $10^7$m/s, which is of the order the vacuum-speed of light $c$. That consideration is examined in more detail in Section 3 of this work.

The mass of the dominant particle $d$ of galaxies, which is expected to be also the dominant particle of the observable universe, must be given by $m_g(D_{dg})$, where $D_{dg}$ is the fractal dimension relating the parameters of the particle to the galactic parameters. In the case of the CMSI model, $D_{dg}=2$, but the dimension is presumed to be unspecified for the present investigation. Figure 1 presents the logarithm of the mass $m_g(D)$ in (2.1) for the independent variable $D$ ranging from 1 to 3. The term $m_g(D)$ ranges from $10^{27}$eV, which is of order near the Planck mass, to a few eV, for $D=1$ and $D=3$, respectively. If the fractal dimension $D_g$ relating the galactic parameters to the parameters of the dominant particle is equal to 2 then the mass of the dominant particle must be of the order $m_g(2) \sim 1$GeV. That conclusion is significant in the context of the dark matter problem since it indicates that the mass of the putative dark matter particle must be of order near the nucleon mass in a fractal hierarchy whose dimension is 2 and also in the CMSI model.

*3. The mass of the dominant particle from quantum mechanics*

The characteristic radius $R_x$, defined in the context of (1.7) – (1.10), that is associated with any fundamental particle $x$ must be no less than the Compton wavelength $l_x=h/(M_x c)$. It follows from (1.8) and $R_x > l_x$ that the mass $M_x$ of the dominant particle of some body $i$ must be no less than $\mu_h(D_{xi})$, where

$$\mu_h(D) \equiv \left( \frac{q_D^{1/D} h}{c} \right)^{\frac{D}{D+1}} \tag{3.1}$$

for some unspecified dimension $D$. The term $\mu_h(D)$ represents a quantum-mechanical lower bound on the mass of the dominant particle that could be consistent with the scaling laws (1.7) – (1.10). For $D=2$ the requirement in (3.1) provides also the lower bound on the mass of a particle that could be consistent with the CMSI model.

The parameter $q_D$ should be of the order $M_g/R_g^D$ and may be estimated from the characteristic galactic parameters presented in Section 2. In Figure 1 is presented the mass $\mu_h(D)$ for $D$ ranging from 1 to 3, with $q_D$ obtained from the estimated parameters of a typical galaxy. For all dimensions $D$ less than approximately 2, the mass $m_g(D)$ of the dominant particle in galaxies is smaller than the lower bound $\mu_h(D)$. Therefore, the dimension $D_{dg}$ must not be less than approximately 2, and that the mass of the dominant particle $d$ of galaxies, and presumably also of the observable universe, could not be less than of the order the mass of the nucleon. It is noteworthy that the fractal dimension 2 and the scale of mass of the nucleon mass, which represents the CMSI model, corresponds to the minimum particle mass for any fractal dimension.

It follows from (1.10) and $A_x \sim \hbar$ that the characteristic radius $R_x$ of the dominant particle $x$ associated with any body $i$ must be given by

$$R_x \sim \frac{\hbar}{M_x f} \sim l_x \frac{c}{2\pi f}. \tag{3.2}$$

In order to be consistent with the basic principles of quantum mechanics, it follows from (3.2) that the constant $f$ must be limited according to

$$f \leq \frac{c}{2\pi}. \tag{3.3}$$

In order to be consistent with the observed parameters of galaxies the constant $f$ must be of the order $A_g/(M_g R_g)$, which is of the order near $c$. It follows from (3.2) and $f \sim c$ that the radius $R_x$ must be of the order $l_x$. The mass $M_x$ must be therefore of the order $\mu_h(D_{xi})$ since the lower bound $\mu_h(D)$ in (3.1) corresponds to the mass for which $R_x \sim l_x$. Since the mass $M_x$ of the dominant particle constituting some body $i$ is, according to (2.1), given by $m_i(D_{xi})$, it follows that

$$m_i(D_{xi}) \sim \mu_h(D_{xi}). \tag{3.4}$$

for any dominant particle $x$ and body $i$. (The relationship in (3.4) reduces to $f \sim c$.) It follows from (3.4) that the dimension $D_{dg}$ relating the parameters of the dominant galactic particle $d$ to the to the characteristic galactic must satisfy $\mu_h(D_{dg}) \sim m_g(D_{dg})$. It is evident from Figure 1 that the dimension $D_{dg}$ must be near 2, and the mass of the dominant particle in galaxies must be of the order the nucleon mass.

*4. The mass of the dominant particle from the cosmological constant*

Let there be some characteristic radius $R_u$ and mass $M_u$ that represent well the parameters of the observable universe. The action $A_u$ should be given by $U_u t_u$, where $U_u$ is the magnitude of the characteristic gravitational potential energy of the observable universe and $t_u$ is the characteristic relaxation time. The characteristic time $t_u$ is of the order $(R_u/g_u)^{1/2}$, where $g_u \sim GM_u R_u^{-2}$ is the magnitude of the characteristic gravitational field, and $G$ is the Newtonian gravitational coupling. The time $t_u$ is therefore of the order $(R_u^3/GM_u)^{1/2}$. The term $U_u$ is given by $U_u \sim GM_u^2 R_u^{-1}$. The characteristic action $A_u$ of the observable universe is therefore of the $(GM_u^3 R_u)^{1/2}$. [1]

The characteristic parameters of the cosmos evolve over time, but there may exist a fundamental upper bound on both $M_u$ and $R_u$. Observations indicate that there exists a vacuum-energy throughout the universe that has caused accelerated cosmic expansion in a manner that is consistent with a positive cosmological constant, $\Lambda$, of the kind first proposed by Einstein. If there exists a cosmological constant then there exists a finite event horizon that approaches asymptotically the De Sitter horizon $R_\Lambda = c\sqrt{3/\Lambda}$. The event horizon $R_e(T)$ is of the order $R_\Lambda$ when the proper time $T$, or "age", of the universe is of the order the fundamental time $T_\Lambda \equiv \Lambda^{-1/2}$, and for all times after. The fundamental cosmic mass $M_\Lambda \sim c^3/(G\Lambda^{1/2})$ represents the largest possible mass $M_e(T)$ that could be contained within the sphere whose radius is $R_e(T)$. The mass $M_e(T)$ is of the order $M_\Lambda$ when $T \sim T_\Lambda$ and for all earlier times. The greatest possible action that could be associated with the contents of the sphere whose radius is the event horizon is of the order [4]

$$A_\Lambda \equiv \left(GM_\Lambda^3 R_\Lambda\right)^{1/2} \sim \frac{c^5}{G\Lambda}. \tag{4.1}$$

The event horizon $R_e(T)$ represents the distance between some point $A$ and the most remote source of particles that could ever be detected at $A$ at any proper time greater than or equal to $T$. The event horizon represents therefore a physically significant measure of the size of the observable universe and could define the characteristic radius $R_u$ and thus also $M_u$.

According to the CMSI model, the characteristic parameters of the observable universe should be determined, roughly, by the stochastic fluctuations of its dominant particle, denoted by $z$ [1]. However, since the cosmic parameters evolve over time, the parameters of the observable universe could not be satisfied at all times. Suppose that $R_u$ and $M_u$ are defined by the event horizon, and there exists some age, represented by the proper time $T_{zu}$, in which the parameters of the universe are related to the parameters of the dominant particle $z$ in a manner that is consistent with the scaling laws in (1.7) – (1.10) for some dimension $D=D_{zu}$. The dimension $D_{zu}$ is treated as being, in principle, distinct from $D_{dg}$, in order the represent the most general case in which the predominant structures may be arranged in a hierarchy that is characterized by a set of fractal dimensions that vary with scale [2].

It follows from (2.1) that the mass $M_z$ of the dominant particle in the observable universe must be given by $m_u(D_{zu})$, where $R_u \sim R_e(T_{zu})$ and $M_u \sim M_e(T_{zu})$. It follows from (1.10) and $f \sim c$ that $A_u$ must be given by $M_u R_u c$ in the epoch when $T \sim T_{zu}$. However, if $A_u$ is given by $(GM_u^3 R_u)^{1/2}$ then, in order for $A_u$ to be of the order $M_u R_u c$, it must be that $M_u \sim c^2 R_u/G$. Since $M_\Lambda \sim c^2 R_\Lambda/G$, and since $R_e(T)$ only increases while $M_e(T)$ only decreases with increasing $T$, there is only one epoch in which $M_u \sim c^2 R_u/G$, which is when $T \sim T_\Lambda$, which is the current epoch [4]. That conclusion is consistent with the analysis in Ref. [1] indicating that the current parameters of the cosmos are consistent with the CMSI model.

If $T_{zu} \sim T_\Lambda$ then the action $A_u$ must be of the order $A_\Lambda$ in (4.1). The mass $M_z$ of the dominant particle in the observable universe must be therefore given by $m_\Lambda(D_\Lambda)$, where

$$m_\Lambda(D) = M_\Lambda \left(\frac{c^5}{G\hbar\Lambda}\right)^{\frac{-D}{D+1}}, \tag{4.2}$$

and the dimension $D_\Lambda$ represents some fundamental, cosmological fractal dimension relating the fundamental mass $m_\Lambda(D_\Lambda)$ of the dominant particle to the cosmic parameters that follow from the cosmological constant observable. It is instructive to note that the term in parentheses on the right side of (4.2) is, aside from a geometrical coefficient, the maximum number $N_\Lambda$ of bits of information that could be registered in a universe with a positive cosmological constant [4]. That fundamental pure number is of the order $10^{122}$ [4]. Associated with the mass $m_\Lambda(D_\Lambda)$ is the number $n_\Lambda(D_\Lambda)$, where

$$n_\Lambda(D) \equiv \frac{M_\Lambda}{m_\Lambda(D)} \sim N_\Lambda^{D/(D+1)}. \quad (4.3)$$

The term $n_\Lambda(D)$ represents the maximum number of particles of mass $m_\Lambda(D)$ that could ever be contained within the sphere whose radius is the event horizon, for some fractal dimension $D$.

It follows from the analysis of Section 3 that the dimension $D_\Lambda$ that determines the mass $M_z \sim m_\Lambda(D_\Lambda)$ of the dominant particle of the observable universe must be the dimension for which $m_\Lambda(D_\Lambda) \sim \mu_h(D_\Lambda)$. Similarly, the dimension $D_{dg}$ relating the parameters of the dominant galactic particle $d$ to the characteristic galactic parameters is such that for which $\mu_h(D_{dg}) \sim m_g(D_{dg})$. The dominant particle $d$ of galaxies is presumably identical to the dominant particle $z$ of the observable universe, and the dimension $D_\Lambda$ must be therefore equal to $D_{dg}$, and there should be one dimension $D_0$ such that $\mu_h(D_0) \sim m_g(D_0) \sim m_\Lambda(D_0)$.

In Figure 1 is presented, along with $\mu_h(D)$ and $m_g(D)$, a plot of $m_\Lambda(D)$ for $D$ ranging from 1 to 3. The masses $m_\Lambda(D)$, $\mu_h(D)$ and $m_g(D)$ all nearly coincide at a certain dimension, which is consistent with the requirement that the particle $z$ be identical to the galactic particle $d$. The dimension $D_0 \sim D_{dg} \sim D_\Lambda$ is approximately 2, and the mass $M_z$ of the dominant particle in the cosmos must be of order near the nucleon mass. The CMSI model, which is equivalent to the fractal scaling laws (1.7) – (1.10) with $D=2$, is thus favored.

Furthermore, if the mass of the dominant particle in galaxies and the observable universe is given by $m_\Lambda(2)$ then it follows from (4.2) that $M_z \sim M_d$ must be scaled to the cosmological constant according to

$$M_z \sim M_d \sim \left(\frac{\hbar^4 \Lambda}{G^2 c^2}\right)^{1/6}. \quad (4.4)$$

The corresponding fractal particle number $n_\Lambda(2)$ is of the order $10^{80}$. That pure number, famously pondered by Eddington, is of order near the baryon number of the observable universe. The fact that the nucleon mass $M_n$ is also of order near $m_\Lambda(2)$ is expected not to be a coincidence. Zel'dovich first proposed that $\Lambda$ should be proportional to $M_n^6$ based on considerations of quantum field theory [5]. Remarkably, holographic principles indicate that $M_n$ is scaled to the cosmological constant in a manner identical to $M_n \sim m_\Lambda(2)$ [6]. The putative relationship $M_n \sim m_\Lambda(2)$ would resolve also a number of problematic large-number coincidences among the parameters of nature [4]. Furthermore, $M_n \sim m_\Lambda(2)$ follows from applying the Bekenstein-Hawking bound to a model for the origin of the universe in which three dimensions inflated from the collapse of seven extra dimensions [7].

It is important to note that the possible physical scaling relationship $M_n \sim m_\Lambda(2)$ requires only that $M_z \sim M_d \sim M_n$, and does not exclude the possibility that the dominant particle in galaxies and the observable universe is non-baryonic. It is, however, very likely that the term $m_\Lambda(2)$ represents a fundamental scale of mass for bodies that are subject to either the CMSI model or the scaling laws associated with a fractal hierarchy whose dimension is 2.

There may exist, in principle, other structures in the universe that are composed of other particles and that are arranged in a manner that is consistent with (1.7) – (1.10) some fractal dimensions other than 2. It is therefore useful to consider several other critical dimensions and masses that follow from (4.2).

It is expected that the largest, physically meaningful fractal dimension is equal to the number of large, spatial dimensions. Since the mass in (2.2) decreases with increasing fractal dimension, the smallest physically significant value for the mass of the dominant particle is $m_\Lambda(3)$, which is

$$m_\Lambda(3) = \frac{M_\Lambda}{N_\Lambda^{3/4}} \sim \left(\frac{\hbar^3 \Lambda}{c^3 G}\right)^{1/4}. \tag{4.5}$$

The mass in (4.5) is of the order 1meV. The corresponding particle number $n_\Lambda(3)$ is of the order $10^{90}$. It is conceivable that the population of neutrinos contained within the observable forms structures characterized by a fractal dimension 3, and that the mass $m_\Lambda(3)$ represents the scale of mass of the neutrino.

The mass $m_\Lambda(1)$ associated with a dimension 1 is

$$m_\Lambda(1) = \left(\frac{\hbar c}{G}\right)^{1/2}, \tag{4.6}$$

which is identical to the Planck mass. The mass $m_\Lambda(1)$ is also the only mass that follows from (4.2) that is independent of the cosmological constant. The particle-mass corresponding to a fractal dimension 1.5 is given by

$$\mu(1.5) = \left(\frac{\hbar^6 \Lambda}{G^4}\right)^{1/10}, \tag{4.7}$$

which is of the order $10^{-20}$kg. The mass in (4.7) is significant mathematically in that it is the only mass that follows from (4.2) that is independent of the speed of light.

It is also instructive to consider two mathematical limits associated with (4.2). The smallest meaningful fractal dimension is 0. The mass $m_\Lambda(0)$ is equal to the maximum cosmic mass $M_\Lambda$ and the corresponding number $n_\Lambda(0)$ is 1. Though not physically meaningful *per se*, that observation does demonstrate that (4.2) satisfies a basic requirement of consistency. The mass $m_\Lambda(0)$ is also the only mass that follows from (4.2) that is independent of the Planck constant. As the fractal dimension $D$ approaches infinity, the mass $m_\Lambda(D)$ approaches the limit

$$\lim_{D \to \infty} m_\Lambda(D) = \frac{M_\Lambda}{N_\Lambda} \sim \frac{\hbar \sqrt{\Lambda}}{c^2}, \tag{4.8}$$

which is of the order $10^{-69}$kg. The fundamental term $\hbar\sqrt{\Lambda}/c^2$ is known as the Wesson quantum, and it is the minimum possible quantum of mass in a universe with a positive cosmological constant [8]. It is essentially equal to the mass of the particle whose Compton wavelength is equal to the De Sitter horizon. That asymptotic behavior satisfies another basic criterion for the consistency of the model presented here. Note also that the asymptotic mass in (4.8) is the only mass obtained from (4.2) that is independent of the gravitational coupling *G*.

___________________________

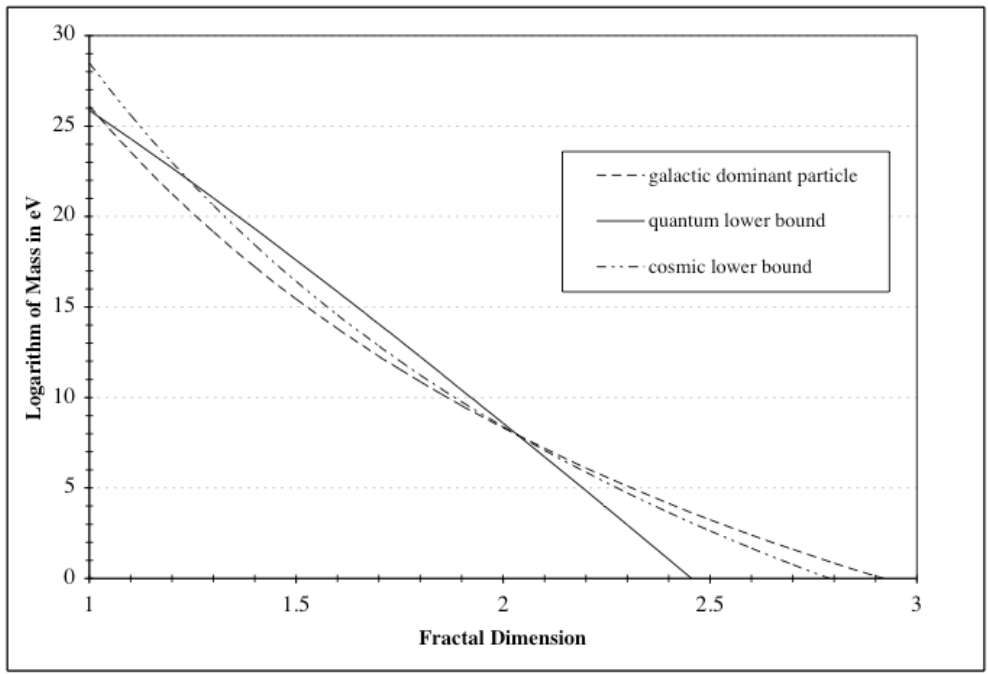

Figure 1. This graph displays the logarithm (base-10) of $m_g(D)$, defined in (2.1), $\mu_h(D)$, defined in (3.1), and $m_A(D)$, defined in (4.2), in electron-Volts (eV), as a function of fractal dimension *D*.

___________________________